\documentclass[11pt,letterpaper]{llncs}
\usepackage{amsmath,amssymb,amsfonts}
\usepackage{graphicx}
\usepackage{marvosym}
\usepackage{url}
\usepackage{fancyhdr}
\usepackage{makecell} 
\usepackage{rotating}
\usepackage{multirow}
\usepackage{enumitem}
\usepackage{cite}

\usepackage{geometry}
\newcommand{\keywords}[1]{\par\addvspace\baselineskip
\noindent\keywordname\enspace\ignorespaces#1}

\fancypagestyle{firstpage}{\fancyhf{}
\setcounter{page}{1}
\fancyhead[C]{\small{International Conference on Network and Communications (NeTCoM 2024)}}
\rfoot{\thepage}
}

\begin{document}


\title{\LARGE{Security, Trust and Privacy challenges in AI-driven 6G Networks}}


%
%
\author{\large{Helena Rifà-Pous \and Victor Garcia-Font \and Carlos Núñez-Gómez \and \\ Julian Salas}}

\institute{\large{Internet Interdisciplinary Institute (IN3) \\ 
Universitat Oberta de Catalunya (UOC) \\
Center for Cybersecurity Research of Catalonia (CYBERCAT) \\ Barcelona, Spain} 
}

%

%


%
%


\maketitle

\thispagestyle{firstpage}

\begin{abstract}
The advent of 6G networks promises unprecedented advancements in wireless communication, offering wider bandwidth and lower latency compared to its predecessors. This article explores the evolving infrastructure of 6G networks, emphasizing the transition towards a more disaggregated structure and the integration of artificial intelligence (AI) technologies. 
Furthermore, it explores the security, trust and privacy challenges and attacks in 6G networks, particularly those related to the use of AI. It presents a classification of network attacks stemming from its AI-centric architecture and explores technologies designed to detect or mitigate these emerging threats. The paper concludes by examining the implications and risks linked to the utilization of AI in ensuring a robust network.

\end{abstract}


\section{Introduction}
6G is the next phase in the evolution of wireless communication technologies, succeeding 5G. Compared to its predecessors, 6G will utilize higher frequency bands enabling wider bandwidth and lower latency. This higher frequency and bandwidth are expected to result in faster, more stable, and more reliable network connections that will support a wide range of applications and services, enhancing 5G already enabled transformative experiences such as virtual reality (VR) and augmented reality (AR), and enabling new, innovative experiences like immersive extended reality (XR), massive IoT connectivity, and digital twins (DT). 

6G will require massive amounts of computing resources to make distributed, complex, and coordinated decisions throughout the whole infrastructure, and so, a highly heterogeneous and distributed infrastructure is needed. Cellular networks are traditionally organized into three main parts: the Radio Access Network (RAN), the Transport Network (TN), and the Core Network (CN). RAN serves as the vital interface that links user devices to the cellular network, granting access to diverse network services. It comprises base stations and other radio access nodes. TN facilitates seamless data transfer between RAN and CN, managing both user and control plane traffic. CN assumes the role of the central intelligence hub, manages user sessions, mobility, and other network functions, and is responsible for connecting to external networks, such as the internet and other cellular networks.

In the context of 6G, these networks are expected to evolve towards a more disaggregated network structure and the boundary between the access network and core network will blur. The roles of RAN, TN, and CN will become more intertwined facilitating the enrichment of interactions between the three domains. In particular, functions that were traditionally confined to one domain can now be colocated, enhancing efficiency and communication. Moreover, the concept of merging similar functions emerges, eliminating redundancy and streamlining the network's capabilities. This shift is instrumental in achieving the network flexibility and scalability requirements of 6G.

The evolution of CN focuses on network softwarization to enhance efficiency, flexibility, scalability, and performance. This transformation relies on key technologies like Software Defined Networking (SDN) and Network Function Virtualization (NFV). 

SDN introduces a separation of the control and data planes within network architectures, enabling the abstraction, programming, and management of network infrastructure through software-defined functions. It also assumes a central role in the orchestration and management of network slicing, which is a transformative paradigm that allows the logical partitioning of the CN into distinct and dedicated virtual networks or slices. Each network slice is tailored to meet the specific requirements of designated services, enabling multiple services and applications to coexist harmoniously on a shared physical infrastructure.

Each layer of the CN comprises a set of Network Functions (NFs). Each NF follows a two-level service structure, with microservices acting as fine-grained components that combine to create more comprehensive NFs. This flexibility in service granularity enables the 6G CN to cater to a wide range of communication needs across various application scenarios. NFV abstracts NFs from dedicated hardware appliances, rendering them as software-based instances. This virtualization injects a heightened degree of flexibility and efficiency into the allocation of network resources, aligning perfectly with the dynamic nature of 6G services.

On the other hand, the evolution of RAN is characterized by multivendor, interoperable components that can be programmatically optimized through a centralized abstraction layer and data-driven closed-loop control \cite{niknam2022intelligent}. 6G RAN architectures are founded on disaggregated, virtualized, and software-based components, linked via open and standardized interfaces, ensuring interoperability among different vendors. This disaggregation strategy aligns with cloud-native principles, which enhance network resiliency and reconfigurability.

Beyond disaggregation, the second main innovation of RAN is that it introduces an open architecture (O-RAN) based on RAN Intelligent Controllers (RICs), which introduce programmable components designed to optimize and orchestrate the network through data-driven closed-loop control. Equipped with AI tools and models, RICs handle tasks such as traffic steering, interference management, QoS management and load balancing. Additionally, RICs support third-party applications (rApps and xApps) designed to enhance RAN optimization, including policy guidance, configuration management, and data analytics. 

The third innovation of the RAN architecture is the incorporation of additional components for managing and optimizing network infrastructure, including edge systems and virtualization cloud platforms. Virtualization within RAN networks is expected to reduce power consumption by dynamically scaling compute resources according to user requirements, minimizing energy use to only what is necessary \cite{polese2023understanding}. 

Both RAN and CN evolve towards a fog-edge-cloud continuum architecture. Fog-edge-cloud continuum represents a spectrum or a continuum of cloud resources and services that extends from centralized cloud data centers to the network's periphery in a distributed and fog-edge computing infrastructure. This approach involves deploying servers closer to end devices, thus reducing traffic overhead and ensuring stringent Quality of Service (QoS). Besides improving performance, this approach can also mitigate information leakage risks and data tampering threats, as data is transmitted to adjacent edge servers instead of remote cloud servers.  The integration of edge computing, Multi-Access Edge Computing (MEC), and the fog-edge-cloud continuum is poised to play a pivotal role in 6G \cite{letaief2021edge,oztoprak2023technological}. 

On an another hand, 6G networks usually incorporate Internet of Things (IoT) nodes, which assume critical roles in essential infrastructures.  These devices facilitate real-time data collection, analysis, and decision-making, enhancing operational efficiency and enabling innovative services. 

One key technology to achieve the integration of these multiple components of a 6G network is the artificial intelligence (AI) \cite{letaief2021edge,oztoprak2023technological}, which enables adaptive radios and autonomous network management, thus enhancing network performance, resource optimization, security, and customization of services \cite{AHAMMED2023197}.

AI will play a pivotal role in different aspects of 6G networks. In the context of the radio interface, intelligent radios that integrate AI capabilities into radiofrequency technologies will be used \cite{letaief2019roadmap}. This integration will enable radios to become more adaptive, perceptive, and self-aware, allowing them to sense, understand, and respond to changes in the radio frequency (RF) spectrum and supersee situations of different wireless problems, like fading, interference, etc.

Besides, AI technologies will be employed to enhance the whole end-to-end communications \cite{klaine2017survey}. An AI algorithm can be used to train the transmitter, channel, and receiver as an auto-encoder, so that the transmitter and receiver can be jointly optimized. 6G systems are very complex and traditional rule based optimization methodologies are very hard to apply. The use of AI solutions reduces the complexity associated with network management and optimization. It can use the feedback loop between the decision maker and the physical system to automatize management decisions and iteratively refine their actions in a closed loop based on the system’s feedback to eventually reach optimality. 

AI also plays a crucial role in making autonomous 6G networks \cite{zhang2020towards}. With built-in AI engines, the 6G system can automatically organize the network structure and manage various resources like slices, computing power, caching, energy, and communication. This enables the system to adapt smoothly to changing demands. AI-based topology and resource management are essential for efficiently adjusting resource usage based on dynamic user needs and evolving environmental conditions.

6G networks will hold massive nodes and data, paving the way for mobile data analytics \cite{zhang2019deep}. This will enable predicting user behavior and environmental circumstances, resulting in more personalized services, efficient resource use, and support for emerging technologies like smart cities and virtual reality.

Finally, AI will be used to improve network security, trust, and privacy of the infrastructure, software, and final nodes \cite{9482503}. It can enhance network security by detecting and preventing cyberattacks, analyzing network traffic patterns, safeguarding sensitive information and identifying suspicious activity. It can also automate, enhance, or complement various network security functions, such as response and recovery.

This article offers a comprehensive overview of the cybersecurity challenges present in 6G networks, addressing security, trust, and privacy concerns. It particularly delves into AI-related issues, recognizing AI as a key distinction from previous mobile network generations, and presents a classification of network attacks stemming from its AI-centric architecture. Moreover, it explores technologies designed to detect or mitigate these emerging threats.

Existing literature \cite{veith2023trust,Mao:2023,9482503,abdel2022security} examines general network attacks in 6G networks without exploring the implications and risks linked to the utilization of AI in ensuring a robust network across the three crucial dimensions: security, trust, and privacy.


\section{Security challenges and attacks}
\label{sec:security}
The security landscape of 6G networks is shaped by several critical factors. Next, we will examine the main challenges and attacks faced by 6G networks.

\subsection{Security challenges in 6G}
The security challenges of 6G networks are driven by the fog-edge-cloud continuum architecture of the network, the increasing softwarization and virtualization, the use of interoperable multivendor components and IoT devices, and the massive connectivity and mobility inherent in 6G, which expand the threat surface \cite{wang2020security, abdel2022security}. Following we outline the main ones:

\begin{itemize}
    \item \textbf{Physical tampering}: The 6G network is highly vulnerable to physical tampering of nodes. Deploying MEC at the edge with lightweight devices compromises both the integrity of the devices and the data they process. Moreover, relying on IoT systems to gather data from critical infrastructures exposes networks to threats like resource exhaustion, insecure communication, and physical intrusion due to their limited security controls and computational capacity.

    \item \textbf{OS and Protocol Heterogeneities}: The heterogeneous cloud/edge infrastructure in 6G networks lacks standardization, presenting interoperability challenges for security mechanisms. This complexity may result in disparate nodes utilizing outdated algorithms or inadequate key lengths, hindering secure communication agreements.

    \item \textbf{User interfaces}: Limited user interface in many end devices (IoT, gadgets), hampers threat awareness and response.
    
    \item \textbf{Weak computation power}: Edge devices at the periphery lack robust defense mechanisms.   

    \item \textbf{Security protocols}: The computational demands of traditional security solutions may not meet the efficiency required for 6G networks, particularly in low-latency, high-speed communications.

    \item \textbf{MEC containerization}: MEC utilizes containerization for seamless integration into current environments. However, in resource-constrained setups, containers may run on compromised hosts or dishonestly consume significant resources, incapacitating other containers.
    
    \item \textbf{SDN constraints}: The decoupling of the control and data planes in SDN introduces vulnerabilities that attackers can exploit. SDN switches, with limited memory, are vulnerable to resource saturation attacks, while the centralized controller is a prime target for malicious actors, potentially compromising network performance and integrity. Moreover, SDN switches heavily rely on the controller for decision-making, posing the risk of overloading and causing network disruptions.
    
    \item \textbf{Complex security enforcement}: Security in edge computing is complicated due to the frequent mobility of network entities across different administrative domains. Besides, it requires fine granularity in access control that current models often overlook. 
     
    \item \textbf{Open interfaces}: RAN promotes the use of open interfaces, which expands the attack surface and exposes the system to third-party code.
     
    \item \textbf{Multi-vendor interoperability}: The involvement of multiple vendors and service providers can lead to inconsistent security implementation, patch management processes, and varying levels of security expertise among them.

    \item \textbf{Powerful attacks}: Adversaries, including AI-driven entities, possess significant power and have a very large attack surface, capable of orchestrating intelligent, widespread, and prevalent attacks.    
\end{itemize}

Table \ref{tab:components_security_challenges} provides a summary of the security challenges and identifies their associations with the architecture 6G features. Challenges linked to edge/cloud are those where security vulnerabilities stem from the execution of network functions, both in the RAN or in the CN. 

\begin{table}[!htbp]
\centering
\caption{Components related to security challenges}
\label{tab:components_security_challenges}
\resizebox{\columnwidth}{!}{%
\begin{tabular}{|p{0.26\linewidth}|p{0.13\linewidth}|p{0.13\linewidth}|p{0.13\linewidth}|p{0.13\linewidth}|p{0.13\linewidth}|}
\hline 
 \textbf{Challenges} & IoT devices & Edge/Cloud & Virtualizat. & Multivendor & Massive \\
\hline 
 \textbf{Physical tampering} &x  & x &  &  &  \\
\hline 
 \textbf{Heterogeneities} & x &  &  &  & x \\
\hline 
 \textbf{User interface} & x &  &  &  &  \\
 \hline 
 \textbf{Computational power} &  & x &  &  &  \\
\hline 
 \textbf{Security protocols} &  & x &  &  &  \\
\hline 
 \textbf{Containers} &  & x & x &  &  \\
 \hline 
 \textbf{SDN constraints} &  &  & x &  &  \\
\hline 
 \textbf{Complex sec. enforce.} &  &  & x & x & x \\
\hline 
 \textbf{Open interfaces} &  &  &  & x &  \\
\hline 
 \textbf{Interoperable} &  &  &  & x &  \\
\hline 
 \textbf{Powerful attacks} &  &  &  &  &  x\\
 \hline
\end{tabular}
}
\end{table}

\subsection{Security attacks in 6G} 

Security attacks on AI systems in a 6G network can be classified into: (1) Poisoning attacks, which contaminate the system's training phase, and (2) Evasion attacks, occurring during the operational phase, in which the adversary attempts to evade the system.

In the case of poisoning attacks, they can be divided into three main groups \cite{nguyen2021security}: (1) data poisoning, which involves altering training data or input objects to deceive machine learning algorithms; (2) algorithm poisoning, aiming to influence the distributed learning process by manipulating weights in local learning models; and (3) model poisoning, where the deployed model is replaced with a malicious one.

\textbf{Data poisoning} is the most common of the poisoning attacks and can lead to false predictions and biased decision-making. 
In traditional batch or offline learning models, data poisoning attacks on the training dataset (such as label flipping or data cleaning) occur through data injection attacks, such as SQL injection, that introduce or modify database training data, or data modification attacks, such as impersonation attacks that exploit vulnerabilities in authentication protocols to spoof another user, or privilege escalation attacks to gain higher levels of access than intended. Injection attacks are deterred by encryption, data integrity checks and continuous monitoring for anomalous data patterns, while solutions for modification attacks focus on zero-touch networks where devices are automatically configured, provisioned, managed and maintained, and strong authentication methods and least priviledge principles are implemented.

In AI systems relying on online learning, where models are continually updated with streaming data, vulnerabilities may also arise from Man-in-the-Middle attacks or selective packet dropping attacks, both of which tamper with the data stream used for model training. To mitigate these risks, it's advisable to employ strategies like adopting a zero-trust architecture and implementing intrusion detection systems equipped with anomaly detection capabilities.

\textbf{Algorithm poisoning} attacks typically occur in distributed learning contexts like federated learning. In this case, backdoor attacks may manipulate weights in local models, or a node may send a compromised model update to the central server. DoS attacks can render nodes unavailable during federated updates, introducing bias into the model. Slow DDoS attacks masquerade as legitimate traffic while draining resources and causing service disruption.
To counteract threats, machine learning models can identify malicious users, and blockchain technology can authenticate users. 

\textbf{Model poisoning} attacks involve modifying learning algorithms through logic corruption attacks, often initiated through black-box adversarial attacks. Defensive measures include AI anomaly detection.

\textbf{Evasion attacks} during AI operation involve miss-classification adversarial attacks, where the adversary carefully crafts imperceptible perturbations in the input data, forcing the AI model to misclassify the perturbed samples. Defensive measures include robust data validation, diverse datasets, data authenticity verification, and continuous monitoring.

Table \ref{tab:6g_attacks} provides a summary of security attacks, linked with the 6G security challenges and mitigation strategies.

\begin{table}[!htb]
\centering
\caption{Security attacks, challenges and mitigation strategies in 6G}  
\label{tab:6g_attacks}
\resizebox{\columnwidth}{!}{%
\begin{tabular}{|p{0.03\linewidth}|p{0.11\linewidth}|p{0.23\linewidth}|p{0.22\linewidth}|p{0.25\linewidth}|}
\hline 
 \multicolumn{3}{|c|}{\textbf{Attack}} 
 & \textbf{Challenges}
 & 
 \textbf{Mitigation strategy}
 \\
\hline 
 \multirow{4}{*}{
\rotatebox{90}{Poisoning}}
 & \multirow{2}{*}{\makecell[l]{Data \\ Poisoning}} & \makecell[l]{\textit{In dataset learning:}\\Data injection (SQL)\\Impersonation \\Priviledge escalation} & \makecell[l]{Containers\\SDN construction\\Complex security} & \makecell[l]{Encryption\\Data integrity checks\\Strong authentication\\Least privilege principle\\Anomaly detection} \\
\cline{3-5} 
   &   & \makecell[l]{\textit{In online learning:}\\Man-in-the-middle\\Selective dropping} & \makecell[l]{Physical tamp.\\Heterogeneities\\User interfaces\\Computational pow\\Security protocols\\} & \makecell[l]{Intrustion detection\\Zero-trust architect.} \\
\cline{2-5} 
   & Algorithm Poisoing & \makecell[l]{Backdoors\\DoS (Slow DDoS)} & \makecell[l]{Open Interfaces\\Interoperable} & \makecell[l]{Intrusion detection\\Blockchain} \\
\cline{2-5} 
   & Model Poisoning & \makecell[l]{Corruption\\Black-box adversarial} & \makecell[l]{SDN construction\\Complex security\\Powerful attacks} & Anomaly detection \\
\hline 
\parbox[c]{2cm}{
\rotatebox{90}{Evasion}}
& \multicolumn{2}{l|}{Miss-classification adversarial} & \makecell[l]{Physical tamp.\\Computational pow\\Complex security\\Open Interfaces\\Powerful attacks} & \makecell[l]{Robust data validation\\Use of diverse datasets\\Verify data authenticity\\Anomaly detection} \\
 \hline
\end{tabular}
}        
\end{table}

\section{Trust challenges and attacks}
\label{sec:trust}

Trust plays a pivotal role in facilitating interactions among independent entities, especially in machine-to-machine (M2M) interactions, where it influences connection establishment for data or service exchange.  This aspect is particularly beneficial for enhancing security in distributed networks like 6G where dynamically deployed nodes, belonging to different stakeholders, autonomously establish connections with one another.


\subsection{Challenges for trust models in 6G}
\label{sec:trust_challenges}
The advent of 6G brings forth a landscape characterized by unprecedented connectivity, ultra-low latency and heightened device density levels. Trust in 6G extends beyond conventional security concerns, incorporating dynamic considerations such as distributed architectures and AI-based decision-making approaches. The 6G landscape needs robust reputation mechanisms to establish and maintain trust among network entities. This requires trust frameworks to continuously assess trust values for all participating entities and update these values using information obtained from Quality of Service (QoS) measurements. 
The design of a reputation system can take different approaches. For example, it can be based on (1) a unified set of reputation values that apply universally to all network entities~\cite{li2020bc6g}, or (2) a fully distributed scheme where each entity maintains its own list of reputation values specific to its peers~\cite{sun2007trust}.


Other trust studies have also been carried out on 5G and beyond networks. For instance, Benzaïd et al.~\cite{benzaid2021trust} conducted an analysis of the trust landscape within these networks, identifying key points where trust-related concerns could arise. These concerns span various entities that users must place their trust in, including communication, data, AI/ML models, NFV infrastructure and management and orchestration components. Additionally, the study delves into emerging trust enabler technologies such as blockchain, trusted platforms and behavior analytics.

Integrating AI technologies into 6G also leads to new trust considerations. Goebel et al.~\cite{goebel2018ai} provide insights into the evolution of Explainable AI (XAI) approaches, spanning from early context-bound text-only explanations to more advanced multi-modal methods that offer textual justifications and attention visualization. The trustworthiness of the AI model itself significantly impacts the value of such AI applications.


Below we outline the main trust challenges of 6G networks~\cite{kantola2020trust,kalla2022bc}.
\begin{itemize}
    \item \textbf{Extreme-massive connectivity and sensing.} 6G will bring unprecedented connectivity demands, requiring closer collaboration with third parties and seamless integration with multiple technologies.
    
    \item \textbf{Training misleading.} The reliance on advanced AI/ML models in 6G networks raises the risk of misleading training due to intentional data manipulation by malicious entities, potentially compromising the reliability and trustworthiness of AI systems.

    \item \textbf{Data and infrastructure integrity.} 
    The integrity of data and infrastructure in 6G networks is threatened by compromised software trust chains and interface misconfigurations. Secure software components are critical to prevent the integration of vulnerable elements, while misconfigured interfaces may result in unintended data exposure, unauthorized access, or network instability, eroding system trust.
    
    
    \item \textbf{Trust quantification.} Difficulty to reach an agreement with different parties on unified terms to quantify and evaluate the trust among actors.

    \item \textbf{Trust expansion.} 
    Expanding trust models in 6G networks involves addressing challenges such as geographical diversity, multiple stakeholders, high network speeds, massive number of flow setups, and numerous remote entities. 

    \item \textbf{Emerging technologies.} Creating a distributed trust model primarily relies on emerging technologies such as Distributed Ledger Technology (DLT), which are still in their early stages, leading to ambiguity in technology governance, regulatory uncertainties, and concerns about computational and communication sustainability. Although numerous mitigating strategies exist, persisting security concerns, including 51\% attacks and transaction privacy vulnerabilities, continue to demand attention.
\end{itemize}

Table \ref{tab:components_trust_challenges} offers a concise overview of the trust challenges encountered in 6G networks, elucidating their affiliations with particular network segments.

\begin{table}[!htbp]
\centering
\caption{Components related to trust challenges}
\label{tab:components_trust_challenges}
\resizebox{\columnwidth}{!}{%
\begin{tabular}{|p{0.27\linewidth}|p{0.13\linewidth}|p{0.13\linewidth}|p{0.13\linewidth}|p{0.13\linewidth}|p{0.13\linewidth}|}
\hline 
 \textbf{Challenges} & IoT devices & Edge/Cloud & Virtualizat. & Multivendor & Massive \\
\hline 
 \textbf{Extreme connectivity} &x  &  &  &  & x  \\
\hline 
 \textbf{Training misleading} & x & x &  & x &  \\
\hline 
 \textbf{Integrity}  & x &  & x &  &  \\
 \hline 
 \textbf{Trust quantification} &  &  &  & x & x \\
\hline 
 \textbf{Trust expansion} &  & x &  & x & x \\
\hline 
 \textbf{Emerging technologies} &  & x &  &  &  \\
 \hline 
\end{tabular}
}
\end{table}

\subsection{Attacks against trust models in 6G}
\label{sec:trust_attacks}
In the emerging landscape of 6G networks, ensuring trust becomes paramount. Following we delve into the intricate domain of trust-related attacks~\cite{benzaid2021trust,kang2020federated}. For each identified attack, we outline effective countermeasures to fortify the foundation of trust in 6G environments~\cite{huang2021trust,veith2023trust}. We also provide a concise overview relating the trust challenges previously discussed to the identified attacks and mitigation measures in Table \ref{tab:trust_overview}.

In the scope of AI technologies, \textbf{poisoning attacks} represent a significant attack vector. When this attacks occur in the training phase not only jeopardize the system's security (as discussed in Section \ref{sec:security}) but also erode trust, as the accuracy of results declines, undermining network performance and the confidence in the system. 

Similarly, \textbf{evasion attacks} driven by adversarial confidence-reducing exploits further degrades trust, since the confidence of prediction if very low. The resilience of AI systems can be improved training the models using a diverse set of adversarial examples.

Within the same AI scope, \textbf{exploratory attacks} based on model extraction  attempt to illicitly obtain sensitive information from deployed machine learning models. Adversaries seek to replicate targeted models, exploiting vulnerabilities and internal architecture to execute unauthorized actions, potentially undermining network trust. Mitigation strategies against such attacks involve injecting noise into the execution time of the machine learning model to introduce variability and hinder the precise extraction of model parameters.

Secure identity management also plays an important role in 6G settings. Attacks such as \textbf{identity spoofing} can have a significant detrimental effect on the trust and authenticity of digital identities within a network. In this context, trust is established through the verification of user identities, ensuring that parties communicating or interacting are who they claim to be. Identity spoofing disrupts this trust by impersonating legitimate identities, leading to various security and privacy risks. The adversary creates a false identity or impersonates a general role or device to deceive systems or gain access to resources. Countermeasures are based on multi-factor authentication and strong verification processes.

Additionally, other identity-related attacks can be perpetrated by properly identified users or servers. \textbf{Insider attacks} are security breaches or malicious activities initiated by trusted entities who have privileged access and knowledge of the network's internal operations. Unlike external threats, insider attacks leverage their legitimate positions or in-depth knowledge to exploit vulnerabilities, potentially causing substantial harm to the network's integrity, confidentiality and availability. Mitigation strategies are based on zero-trust models.

In 6G, \textbf{repudiation attacks} can compromise transaction and communication reliability and accountability by entities disavowing prior actions, leading to trust issues. Mitigation involves deploying authentication, authorization, and logging mechanisms to maintain clear audit trails and employing digital signatures and cryptographic methods for non-repudiation.

The adoption of DLT solutions in the 6G landscape can be advantageous in terms of data integrity and action tracking, but it could also introduce new attack vectors. Regarding blockchain and smart contract-based solutions, re-entrancy attacks pose a significant security risk. This threat involves an attacker making repeated calls or re-entering a vulnerable smart contract. This malicious activity exploits the contract's state, leading to unauthorized actions and potentially undermining network trust. Effective mitigation of re-entrancy attacks necessitates rigorous code auditing and comprehensive testing measures to identify and rectify vulnerabilities.

\begin{table}[!htb]
\centering
\caption{Trust attacks, challenges and mitigation strategies in 6G}
\label{tab:trust_overview}
\resizebox{\columnwidth}{!}{%
\begin{tabular}{|p{0.22\linewidth}|p{0.39\linewidth}|p{0.39\linewidth}|}
\hline
\textbf{Attack}   & \textbf{Challenges}                                                  & \textbf{Mitigation strategy}                                                                              \\ \hline
Poisoning         & Training misleading                                                  & Moving target defense and input validation                       \\ \hline
Evasion  & Training misleading                                        & Adversarial examples                                                       \\ \hline
Model extraction  & Data/infrastructure integrity                                        & Noise injection                                                                                           \\ \hline
Identity spoofing & Massive connectivity, Data/infrastructure integrity, Trust expansion & Multi-factor authentication and strong verification processes                                             \\ \hline
Insider attacks   & Massive connectivity, Data/infrastructure integrity, Trust expansion & Zero-trust models                                                                                         \\ \hline
Repudiation       & Trust quantification, Trust expansion, Emerging techs         & Robust authentication, authorization and logging mechanisms. Digital signatures and cryptographic methods \\ \hline
Re-entrancy       & Emerging techs                                                & Code auditing                                                                                             \\ \hline
\end{tabular}%
}
\end{table}

\section{Privacy challenges and attacks}
The privacy framework of 6G networks is influenced by various pivotal factors. Next, we will delve into the primary challenges and vulnerabilities inherent in 6G networks.

\subsection{Privacy challenges in 6G}
In the 6G era, increased data generation, storage, and processing pose notable privacy challenges, incorporating sensitive information like precise location tracking and behavioral predictions. The integration of 6G biosensing, which includes accessing intimate health data, escalates risks such as fraud, blackmail, intrusive marketing, and pervasive surveillance. 
Furthermore, companies struggle to safeguard sensitive data used in access control systems, where threats such as ransomware attacks and corporate espionage jeopardise data integrity. 
%

The complexity of networks and the diversity of applications
in 6G will probably make data privacy preservation more difficult than ever~\cite{Nguyen:2021}.
Federated Learning (FL) and Privacy Enhancing Technologies (PETs) such as differential privacy (DP)~\cite{Dwork:2014}, homomorphic encryption (HE)~\cite{Rivest:1978}, secure multi-party computation (SMC)~\cite{Yao:1982}, are appropriate technologies that will protect personal data, complying with GDPR standards, and satisfying the
needs for statistical models.

Federated Learning addresses the confidentiality
issue by keeping datasets locally, yet the shared model updates are prone to privacy leakage, through model inversion or membership inference attacks.

Differential privacy provides a privacy-preserving mechanism to guarantee a level of privacy disclosure for local datasets by adding random noise \cite{Dwork:2014}. 
The wireless channel noise properties can be used as a
privacy-preserving mechanism, and a DP constraint
using such random noise will cause no performance degradation with respect to a non-private design as long as the signal-to-noise ratio is sufficiently low~\cite{Liu:2020}. 
Secure Multiparty Computation aims to protect a distributed computation model from the inputs of communication parties while keeping
those inputs private.
Homomorphic Encryption allows performing operations, such as search and query, on encrypted data directly without decryption. 

These four key technologies (FL, DP, HE and SMC) are the building blocks to tackle the main privacy challenges of 6G networks~\cite{Ferrag:2018,Liyanage:2022}
listed as follows: 
\begin{itemize}
    \item \textbf{Complex identity management:} The advent of Internet of Everything (IoE) will introduce personal IoT networks, including wearable devices and IoT devices in offices and factories. The growing number of connected devices per person will pose significant identity management challenges.

    \item \textbf{Eavesdroppers:} With IoT applications managing increasingly sensitive data, the threat of data theft by eavesdroppers has grown. IoT devices often lack the resources for efficient encryption, and eavesdroppers, remaining undetected, pose challenges to edge network security.

    \item \textbf{User-generated data:} In the context of edge intelligence, AI models are trained using extensive user-generated data, including sensitive private information, which is accessible to edge servers for model training and execution. The servers can be honest but curious, inferring personal data that can be used for other purposes beyond network traffic optimization.

    \item \textbf{MEC containerization:} Containers are extensively employed in the edge network. They offer several advantages, notably reduced startup time and decreased resource utilization. However, it is crucial to acknowledge that containers do not provide an equivalent level of isolation when compared to Virtual Machines (VMs) used in cloud networks. Containers share access to kernel-based filesystems, thereby posing security challenges. A potentially malicious container could exploit this shared access to gain unauthorized entry into and potentially extract information from other co-hosted containers. 
\end{itemize}

Table \ref{tab:privacy_challenges} provides a summary of the privacy-related challenges within 6G networks, clarifying their associations with the architecture 6G features.

\begin{table}[btp]
\centering
    \caption{Components related to privacy challenges}
\label{tab:privacy_challenges}
\resizebox{\columnwidth}{!}{%
\begin{tabular}{|p{0.2\linewidth}|p{0.13\linewidth}|p{0.13\linewidth}|p{0.13\linewidth}|p{0.13\linewidth}|p{0.13\linewidth}|}
\hline 
 \textbf{Challenges} & IoT devices & Edge/Cloud & Virtualizat. & Multivendor & Massive \\
\hline 
 \textbf{Identity mgt} &  &  &  &  & x \\
\hline 
 \textbf{Eavesdroppers} & x &  &  &  &  \\
\hline 
 \textbf{User-gen data} &  & x & x & x &  \\
 \hline 
 \textbf{Containerizat.} &  &  & x &  &  \\
 \hline
\end{tabular}
}
\end{table}

\subsection{Privacy attacks in 6G}

Privacy attacks in 6G encompass a diverse array of malicious techniques and strategies that threaten the confidentiality, integrity, and availability of sensitive data traversing the network \cite{Ahmad:2023,Mao:2023, Khan:2019}. Our exploration also offers insight into effective mitigation strategies aimed at fortifying the network's defenses. 

\begin{itemize}
    \item \textbf{Deception attacks} are a form of cyberattack in which attackers manipulate or mislead network components, services, or users into making incorrect decisions or revealing sensitive information. In the context of 6G networks, where the integration of emerging technologies like NFV, SDN, and MEC is prevalent, deception attacks pose significant threats. Countermeasures are based on threat intelligence sharing, security audits and technologies to create decoy assets and traps within the network.

\item \textbf{Side-channel attacks} compromise user security by exploiting publicly accessible non-sensitive information, analyzing physical parameters like electromagnetic emissions and execution time. Attackers can manipulate a cache's content or introduce faults, compromising cryptographic operations and extracting secret keys. In network slicing, attackers can indirectly affect other slices, highlighting the need for strong isolation and avoiding hosting similar applications on slices with similar hardware configurations. 
Countermeasures include avoiding hosting applications on slices with similar hardware configurations and ensuring strong isolation among slices.

\item \textbf{Information disclosure} exploits intercepting information by unauthorized entities to compromise system security objectives like user traceability. This may lead to the launch of other attacks. Mitigation strategies are based on robust authentication and authorization protocols.

\item \textbf{Data Leakage} attacks involve the unauthorized or malicious transfer of information or data from one network slice to another. These attacks often target shared Network Functions (NFs) and can undermine data privacy and network security. To mitigate such attacks advanced trusting mechanisms are essential such as trust and reputation models, computational trust, blockchain or zero-trust security.

\item \textbf{Location tracking} attacks exploit vulnerabilities of network slicing-enabled networks to track the location of a target user. These attacks can be performed in several ways, including, exploiting design flaws or misbehaving NFs, and compromising network edge functions. The primary mitigation strategies include continuous NF behavior monitoring and rigorous design audits.

\item \textbf{Eavesdropping} attacks involve malicious interception and monitoring of communication between legitimate parties. These attacks are highly detrimental as they compromise the confidentiality and privacy of transmitted data. To mitigate eavesdropping attacks, several strategies can be employed like end-to-end encryption and intrusion detection systems. 

\item \textbf{File injection} attacks in the context of 6G involve malicious actors injecting unauthorized files or data into the network with the intent of compromising the privacy of the users. These attacks may include injecting malware (e.g., spyware) or malicious code into legitimate data flows, leading to the leakage and exposure of private information. File injection attacks are deterred by employing strong encryption, data integrity checks, and continuous monitoring for anomalous data patterns in network traffic. 

\item \textbf{Model inversion} is a privacy threat where an adversary attempts to recover sensitive information about a user or entity by analyzing the outputs of a machine learning model. This attack specifically aims to reverse-engineer the underlying data that was used to train a model. In 6G networks, where machine learning and AI-driven services are prevalent, this type of attack can have serious privacy implications. An effective defensive approach involves exerting control over the information disclosed through ML APIs, thereby limiting the insights that attackers can gain. Additionally, to further mitigate the risk of model inversion attacks, introducing noise to ML predictions can be employed as a countermeasure. This added noise introduces uncertainty and complexity into the attacker's inference process, thereby enhancing the security and privacy of the ML model.

\item \textbf{Membership inference} attacks in the context of 6G networks are a privacy threat where an adversary attempts to determine whether a particular user or device is part of a targeted dataset or network group. This type of attack typically involves exploiting information leakage from ML or data processing operations where an attacker tries to infer whether a user's data was included in the model's training dataset. This attack can be used to reveal sensitive information about network participants and their activities. Mitigation strategies include improving the anonymization of training data, employing differential privacy mechanisms, and enhancing data access control to prevent unauthorized access to sensitive datasets. Additionally, ensuring that machine learning models do not overfit to individual users data can help mitigate this type of attack.

\end{itemize}

Table \ref{tab:privacy_attacks} provides an overview of privacy attacks, challenges and mitigation strategies in 6G.

\begin{table}[hbt!]
    \centering
    \caption{Privacy attacks, challenges, and mitigation strategies in 6G}
    \label{tab:privacy_attacks}
\resizebox{\columnwidth}{!}{%
\begin{tabular}{|p{0.2\linewidth}|p{0.38\linewidth}|p{0.42\linewidth}|}
        \hline
        \textbf{Attack} & \textbf{Privacy challenges} & \textbf{Mitigation strategy} \\
        \hline
        Deception & Eavesdroppers, MEC containerization & Deception technology, threat intelligence sharing and security audits \\
        \hline
        Side-channel & Eavesdroppers, MEC containerization & Principle of Least Privilege, separating containers into virtual networks based on their sensitivity, strong isolation of the execution container \\
        \hline
        Information disclosure & User-generated data & AI/ML approaches can be used to learn to identify the appropriate placement policies to prevent hackers from gaining access to sensitive information \\
        \hline
        Location tracking & User-generated data, MEC containerization & Continuous NF behavior monitoring and rigorous design audits \\
        \hline
        Data Leakage & Eavesdroppers, User-generated data, MEC containerization & Advanced trusting mechanisms (trust and reputation models, computational trust, blockchain or zero-trust security) \\
        \hline
        Eavesdropping & Eavesdroppers & End-to-end encryption, intrusion detection systems \\
        \hline
        File injection & Complex identity management, User-generated data, MEC containerization & Strong encryption, data integrity checks, continuous monitoring \\
        \hline
        Model inversion & User-generated data, MEC containerization & Noise Addition, Differential Privacy \\
        \hline
        Membership inference & User-generated data, MEC containerization & Anonymization of training data, enhanced data access control \\
        \hline
    \end{tabular}
    }
\end{table}

\section{Conclusions}
\label{sec:conclusions}
The transition to 6G networks represents a revolutionary advancement in communication capabilities, marked by the seamless integration of virtual realms with connected intelligence, enabling applications like multisensory extended reality and wireless brain-computer interactions. This technological leap, offering blazing-fast data rates, ultra-low latency, and unparalleled reliability, is underpinned by the pervasive role of artificial intelligence (AI). However, amid this transformative potential, it also unveils an intricate and expansive threat surface, demanding innovative security solutions. This evolving security landscape must address the challenges of a highly interconnected, heterogeneous network enriched by cloudification.

On one hand, attacks on the AI models governing the functionality of 6G networks are highly dangerous and challenging to detect due to the complexity of transparency and explainability of this technology. On the other hand, malicious users can also exploit AI to launch more sophisticated and targeted attacks and to lower the barrier to entry for cyberattacks by automating tasks that previously required technical expertise, which puts additional pressure on the system.

Regarding trust, the interconnected nature of networks with components from multiple vendors poses a challenge to establishing trust. Establishing trust levels for nodes utilizing AI is often complex and difficult because the results of AI are hard to interpret, leading to a lack of transparency and explainability. Moreover, if they are from third-party entities, establishing trust becomes even more complicated.

Finally, much of the intelligence of 6G networks will come from distributed AI technologies, such as federated learning. Federated learning leverages distributed data from mobile devices to train AI models collaboratively while safeguarding privacy and reducing resource consumption. However, the nature of federated learning introduces privacy and security risks. Interactions among mobile devices in federated learning scenarios can be exploited for cheating attacks, low-quality local model training attacks, and privacy breaches, posing significant privacy concerns.

In this article, we have explored the cybersecurity challenges in 6G networks, emphasizing security, trust, and privacy concerns, and we have provided a classification of network attacks stemming from the AI-centric architecture.

\section*{Acknowledgment}
This work is linked to the project SECURING PID2021-125962OB-C31,
funded by the Spanish Ministry of Science and Innovation. It is also
supported  by  the  Spanish  Ministry  of  Economic Affairs    and    Digital    Transformation    and    the    European    Union    – 
NextGenerationEU,  in  the  framework  of  the  Recovery  Plan,  Transformation and  Resilience  (PRTR), under the Call  UNICO  I+D  5G  2021  (ref.  number  TSI-063000-2021-13– 6GENABLERS-SEC), and the Call INCIBE (ARTEMISA International
Chair of Cybersecurity and DANGER Strategic Project of
Cybersecurity).

 
\bibliographystyle{splncs04}
\bibliography{references}



\section*{Authors}
\noindent {\bf Helena Rifà-Pous} received her PhD with honors from the Universitat Politècnica de Catalunya (UPC) in Telematics. With experience in industry, she currently serves as an associate professor in the Department of Computer Science at the Universitat Oberta de Catalunya (UOC) and coordinates the MSc program in Cybersecurity and Privacy. 
She is a member of the K-riptography and Information Security for Open Networks (KISON) research group at the Internet Interdisciplinary Institute (IN3), and her research focuses on security and privacy protocols, particularly in distributed and wireless networks (e.g., IoT, 6G), as well as cybersecurity solutions. She has authored numerous articles in journals and conferences, and she also contributes as an editor and reviewer for various journals.

\noindent {\bf Victor Garcia-Font} graduated from Universitat Politècnica de Catalunya (UPC) as a computer engineer in
2005. At the beginning of his career he worked developing information systems for the banking sector
and the public administration. In 2014, he started
an industrial doctorate thesis focusing on information
security in smart cities. At the beginning of 2017 he
presented his doctoral thesis at Universitat Oberta de
Catalunya (UOC) about anomaly detection in smart city
wireless sensor networks. Currently, he is a lecturer
at the Faculty of Computer Science, Multimedia and
Telecommunications at UOC, where he is teaching subjects related to information security and computer networks. Moreover, he is a member of the K-riptography and Information Security for Open Networks (KISON) research group. His current research interests are focused on blockchain, cryptocurrencies, information security and smart cities.\\

\noindent {\bf Carlos Núñez-Gómez} graduated from the University of Castilla-La Mancha (UCLM) as a computer engineer in 2016. After several years working as a software developer and project manager, in 2019 he obtained a M.Sc. degree in information and communication security at the Universitat Oberta de Catalunya (UOC). Subsequently, he started a Ph.D. degree in advanced computer technologies at UCLM. He is currently a member of the K-riptography and Information Security for Open Networks (KISON) research group at the Internet Interdisciplinary Institute (IN3) of the UOC. His research interests include distributed systems, blockchain, information security and fog computing environments. \\

\noindent {\bf Julian Salas} obtained the Ph.D. degree in applied mathematics at the Polytechnic University of Catalunya (UPC) with the Cum Laude and international mentions in 2012. Has been a postdoctoral researcher at the Internet Interdisciplinary Institute (IN3) from
Universitat Oberta de Catalunya (UOC), at the CRISES group from the University Rovira i Virgili (URV) and at the Artificial Intelligence Research Institute-National Research Council (IIIA-CSIC). His main research lines are security and privacy in big data, data mining and graph theory.
\\

\end{document}